\documentclass[prd,twocolumn,showpacs,a4paper,superscriptaddress,nofootinbib]{
revtex4}
\usepackage{graphicx,color,xspace,hyperref}
\usepackage{amsmath,amssymb}

\usepackage[textwidth=18cm,textheight=24cm]{geometry}
\begin{document}
%--------------------------------------------------
\setcounter{topnumber}{1}
%--------------------------------------------------
\title{The cosmological constant: A lesson from the effective gravity of topological Weyl media}
%--------------------------------------------------
\author{G. Jannes}
\affiliation{Low Temperature Laboratory, Aalto University, P.O. Box 15100, FI-00076 AALTO, Finland}
%--------------------------------------------------------------
\author{G.E. Volovik}
\affiliation{Low Temperature Laboratory, Aalto University, P.O. Box 15100, FI-00076 AALTO, Finland}
\affiliation{Landau Institute for Theoretical Physics RAS, Kosygina 2,
119334 Moscow, Russia}

\date{\today}
%--------------------------------------------------

%--------------------------------------------------------------
\begin{abstract}
Topological matter with Weyl points, such as superfluid $^3$He-A, provide an explicit example where  there is a direct connection between the properly determined vacuum energy and the cosmological constant of the effective gravity emerging in condensed matter.  This is in contrast to the acoustic gravity emerging in Bose-Einstein condensates, where the ``value of this constant cannot be easily
predicted by just looking at the ground state energy of the microscopic system
from which spacetime and its dynamics should emerge''~\cite{Finazzi2011}. The advantage of topological matter is that the relativistic fermions and gauge bosons emerging near the Weyl point 
obey the same effective metric and thus the effective gravity is more closely related to real gravity.
We study this connection in the bi-metric gravity emerging in $^3$He-A, and its relation to the graviton masses, by comparison with a fully relativistic bi-metric theory of gravity. This shows that the parameter $\lambda$, which in $^3$He-A is the bi-metric generalization of the cosmological constant, coincides with the difference in the proper energy of the vacuum in two states
(the nonequilibrium state without gravity and the equilibrium state in which
gravity emerges) and is on the order of the characteristic Planck energy scale
of the system. Although the cosmological constant  $\lambda$ is huge, the cosmological term 
$T^{\Lambda}_{\mu\nu}$ itself is naturally non-constant and vanishes in the equilibrium vacuum,  as dictated by thermodynamics. This suggests that the equilibrium state of any system including the final state of the Universe is not
gravitating.
\end{abstract}
%-----------------------------------------------------------------------
\pacs{04.90.+e, 95.36.+x, 67.30.H-}
\maketitle

%{\em Introduction --} 
\section{Introduction}
It is well-known that in Einstein's theory of gravitation  the most
natural candidate for the cosmological constant is the vacuum energy, since this would lead to a non-zero value of the vacuum energy-momentum tensor which should be reflected in the Einstein equation. In condensed
matter analogs of gravity, typically only the first part of general relativity (GR) is
reproduced: quasiparticles move in the effective metric produced by an
inhomogeneity of the vacuum state. The second part of GR --- the Einstein equation
describing the dynamics of the metric field --- as a rule is not reproduced. For example, the acoustic metric obeys hydrodynamic equations, which are certainly not diffeomorphism invariant. As a
result, the analog of the cosmological constant may essentially differ from the
vacuum energy, as was found for acoustic gravity in BEC ~\cite{Finazzi2011} (see also Ref.~\cite{Sindoni2011}).  
We discuss here the connection between the vacuum energy and the cosmological
constant in the effective gravity which  emerges in Weyl topological matter on example of
the  axial $p$-wave superfluid $^3$He-A. 

There are many condensed-matter systems which exhibit an effective relativistic
metric for scalar excitations such as sound or surface
waves~\cite{Barcelo:2005fc}. Superfluid $^3$He-A, Weyl semimetals and graphene provide the effective gravity
for fermionic quasiparticles. Here due to the topologically protected Fermi points
in momentum space 
(3+1 Weyl-points in $^3$He-A~\cite{Volovik2003} and Weyl semimetals~\cite{Abrikosov1971,Abrikosov1998,XiangangWan2011,Burkov2011,Aji2011}, 
and 2+1 Dirac points in graphene
\cite{Vozmediano2010,CveticGibbons2012}) the emergence of effective
gravity is accompanied by the appearance of Weyl and Dirac fermions  and effective gauge
fields intimately related to those of the Standard Model of particle physics,  which all couple to the same effective metric near the topological point.
This is a consequence of the topological theorem -- the Atiyah-Bott-Shapiro construction  -- applied to systems with Weyl, Dirac or Majorana points~\cite{Horava2005}. The emergent gravity has recently also been used to study the anomalies in quantum field theories of topological insulators, such as the gravitational anomaly and thermal Hall effect
~\cite{Stone2012}. Ho\v{r}ava quantum gravity with anisotropic scaling (see~\cite{Griffin2012} and references therein) can be simulated in bi-layer graphene, and the 
corresponding  quantum electrodynamics with anisotropic scaling 
emerging in multilayer graphene has been recently discussed in~\cite{KatsnelsonVolovik2012,Zubkov2012}.

 \section{Thermodynamics and the vacuum equation of state}
\label{vac-eq-of-state} 
It is well-known that Einstein himself considered thermodynamics to be ``the only physical theory of universal content''~\cite{einstein}. There are strong indications that this universal scope of thermodynamics indeed includes gravity. A crucial step was the observation that the Einstein equations can be interpreted as the thermodynamic equation of state of spacetime~\cite{jacobson}, while reports of recent work on the relation between gravity and thermodynamics include~\cite{padmanabhan,blhu}. This suggests that thermodynamics could well play a crucial role for the cosmological constant problem and its relation to the vacuum energy as well~\cite{Volovik2006}.

For a Lorentz invariant vacuum, $\rho_{\rm vac}=-P_{\rm vac}$ is the only possible equation of state as a perfect fluid, and so one can immediately see from the thermodynamic Gibbs-Duhem relation 
\begin{equation}\label{gibbs-duhem}
 P=-\epsilon + Ts +\sum_i  \mu_i q_i~~,~~ q_i=\frac{Q_i}{V}
\end{equation}
(with $s$ the specific entropy and the temperature $T=0$ in the vacuum) that the relevant thermodynamic quantity, which plays the role of the vacuum energy is the analog of grand-canonical energy $\rho_{\rm vac}=\epsilon(q_i)  - \sum_i  q_i d\epsilon/dq_i$. Any conserved quantity $Q_i$,
which characterizes the quantum vacuum,  should be
explicitly taken into account together with its corresponding Lagrange
multiplier $\mu_i$. And indeed it is demonstrated that the quantity which enters the cosmological term in Einstein equations is the density of the grand-canonical energy, $\rho_{\rm vac}$, rather than the energy density $\epsilon$ 
\cite{KlinkhamerVolovik2008,KlinkhamerVolovik2008b,KlinkhamerVolovik2011a}.

By reversing the above argument, one sees that the vacuum equation of state $\rho_{\rm vac}=-P_{\rm
vac}$ is more generally valid~\cite{Volovik2006}. The energy of the vacuum of quantum fields emerging
in a many body condensed matter system is the grand canonical
energy $\rho_{\rm vac}=\epsilon(n)  - nd\epsilon/dn$, where the particle
density $N=nV$ is a conserved quantity and the corresponding Lagrange
multiplier is the chemical potential $\mu=d\epsilon/dn$. The use of the
grand canonical energy here corresponds to the fact that the  many-body
Hamiltonian in second quantization is $\hat H_\text{QFT}=\hat H - \mu \hat N$,
where $\hat H$ is obtained from the Schr\"odinger many-body Hamiltonian and
$\hat N$ is the number operator. The cosmological equation of state  $w= P/\rho$ for the vacuum energy is then again $w=-1$  due to the Gibbs-Duhem relation, regardless of the Lorentz invariance or not of the vacuum.

\section{Bi-metric gravity in systems with Weyl points}
%{\em Bi-metric gravity in systems with Weyl points --}
%In this class of non-relativistic systems, although the effective gravity does not obey the Einstein equation, under certain conditions the analog of the cosmological constant may coincide with the proper vacuum energy, which obeys the vacuum equation of state $\rho_{\rm vac}=-P_{\rm vac}$. The latter is the general equation of state for a vacuum, Lorentz invariant or not, due to the thermodynamic Gibbs-Duhem relation~\cite{Volovik2006}.

Gravity in superfluid  $^3$He-A is bi-metric, with $g_{\mu\nu}$ the effective dynamical
gravitational field acting on the quasiparticles (the Weyl
fermions) and $g^{(0)}_{\mu\nu}$ its value in the
equilibrium vacuum state. The effective spacetime emerging for quasiparticles
is anisotropic due to the uni-axial anisotropy
of the broken symmetry superfluid state:
 \begin{equation} 
g^{(0)\mu\nu}={\rm diag} \left(-1,c_{\perp 0}^2,c_{\perp
0}^2,c_\parallel^2\right)~~,~~\sqrt{-g^{(0)}} = \frac{1}{c_{\perp
0}^2c_\parallel} \,.
\label{background}
\end{equation}
Here $c_{\perp 0}$ is the ``speed of light''
in the direction perpendicular to the anisotropy axis  (the limiting speed
for ``relativistic'' quasiparticles propagating in that direction), and appears as the amplitude
of the equilibrium order parameter in the symmetry breaking phase transition from the normal
liquid $^3$He to  $^3$He-A.
$c_\parallel$ is the ``speed of light'' propagating along the anisotropy axis. One may say that in
$^3$He-A gravity appears as a result of the broken symmetry. Note that $c_\parallel$ is already present in the normal state of liquid $^3$He. In the weak-coupling BCS theory one has $c_{\perp 0}/c_\parallel \ll 1$ with $c_{\perp 0}/c_\parallel\sim 10^{-3}$ in $^3$He-A. Therefore $c_\parallel$ is completely fixed in the low-energy corner where the relativistic quasiparticles emerge. Also note that this anisotropy cannot be detected by internal observers who only dispose of such relativistic quasiparticles to construct rods and clocks with which to
effect measurements~\cite{Volovik2003,Liberati:2001sd,Barcelo:2007iu}. 

We stress that the same effective metric is relevant both for the relativistic fermions emerging in the vicinity of the Weyl  points and for the gauge bosons, which also emerge in the vicinity of the Weyl  points. Thus the effective gravity in $^3$He-A couples universally and interacts with all matter in the same way. This is ultimately a consequence of the Attiyah-Bott-Shapiro topological theorem~\cite{atiyah}. This construction guarantees that the low-energy fermionic modes close to the manifold
of zeroes in their energy spectrum (the generalized Fermi surface) exhibit relativistic invariance in the directions transverse to the surface~\cite{Horava2005}. Lorentz invariance in fermionic systems is therefore obtained in all directions in case the spectrum is characterised by a topologically protected point node: the Fermi point  (Weyl, Dirac or Majorana point)~\cite{Horava2005,Froggatt:1991ft,Volovik:2006gt}.  The effective spacetime emerges as a consequence of the topological universality class, not of its specific realisation, and it is therefore independent of the concrete background metric. Thus, in spite of the (non-relativistic) background structure (the Helium atoms), the effective theory for the quasi-matter obeys both background independence (at least, at the kinematical level\footnote{ Evidently, as remarked above, the dynamical equations of the effective metric are not covariant. However, the energy-momentum tensor in the effective theory nevertheless satisfies a generalized conservation law which accounts for the interaction between vacuum and quasiparticles~\cite{Volovik2003}.}) and the equivalence principle. Both are broken at the trans-Planckian level, as occurs in many scenarios for quantum gravity.

The  grand canonical energy  which governs the transition from normal liquid $^3$He to superfluid $^3$He-A
represents the
analogue of the Ginzburg-Landau energy, and can be obtained from the
$p$-wave pairing model in BCS theory. Let us first consider the perturbations of the metric
which preserve the structure of the equilibrium metric and correspond to a variation of $c_{\perp}$ only, i.e. 
we consider a non-equilibrium metric of the type 
$g^{\mu\nu}={\rm diag} \left(-1,c_{\perp}^2,c_{\perp}^2,c_\parallel^2\right)$.
Then according to Ref.~\cite{Volovik2003}  one obtains that the grand canonical energy of $^3$He-A
at zero external pressure is:
\begin{align}
\rho_{\rm vac}
&=\lambda  \sqrt{-g^{(0)}} \left[ \frac{\sqrt{-g^{(0)}}}{\sqrt{-g}} \ln 
\frac{\sqrt{-g^{(0)}}}{\sqrt{-g}}  
-  \frac{\sqrt{-g^{(0)}}}{\sqrt{-g}}+1\right]~,
\label{VacuumEnergyNon-equilibrium2}
\end{align}
where $\lambda=\Delta_0^4/(6\pi^2\hbar^3)$ with $\Delta_0 =p_F c_{\perp 0}$ the amplitude of the angular dependent superfluid gap
in anisotropic $^3$He-A, and $p_F$ the
Fermi momentum.  $\Delta_0$ plays the analogue role of a Planck energy $E_\text{Planck}$, i.e. the ``UV cut-off'' for the superfluid phase in which the effective gravity emerges.
Close to equilibrium this vacuum energy density is quadratic in deviations from
equilibrium:
\begin{equation}
\rho_{\rm vac}
\approx\frac{\lambda}{2\sqrt{-g^{(0)}}}  \left(
\sqrt{-g}   -  \sqrt{-g^{(0)}} \right)^2~.
\label{VacuumEnergyNon-equilibrium3}
\end{equation}
This demonstrates that in equilibrium, when $g^{\mu\nu}=g^{(0)\mu\nu}$,  the proper vacuum energy 
$\rho_{\rm vac}$ is nullified. This is a consequence of the fact that $^3$He-A
belongs to the class of  self-sustained systems  such as liquids and solids, which may exist
without external environment, and thus at zero pressure. The self-sustained systems 
adjust themselves to external conditions, so that their grand canonical energy is
automatically nullified in the absence of environment 
\cite{KlinkhamerVolovik2008,KlinkhamerVolovik2008b,KlinkhamerVolovik2011a}.  Indeed, $\rho_{\rm
vac}=-P_{\rm vac}$, so for the equilibrium vacuum ($T=0$ and no quasiparticles), one
obtains $\rho_{\rm vac}=P_{\rm ext}=0$.
This nullification is obtained from the macroscopic laws of
thermodynamics. At the microscopic level, it entails a dynamic self-adjustment of the degrees of freedom around the Planck scale. This phenomenon is so well known in condensed-matter physics that the
nullification of $\rho_{\rm vac}=\epsilon-\mu n$ in self-sustained systems in
equilibrium is generally used as a sanity check for numerical calculations of
$\epsilon$ and $\mu$ from the microscopic degrees of freedom.

Since the effective gravity in $^3$He-A does not obey the Einstein equations, there is a priori no reason to define a cosmological term 
$T^{\Lambda}_{\mu\nu}=(2/\sqrt{-g}) \partial \rho_{\rm vac}/\partial g^{\mu\nu}$. But in any case, whether the action is generally covariant or not, note that any (generalized) cosmological term $T^{\Lambda}_{\mu\nu} \propto \partial \rho_{\rm vac}/\partial g^{\mu\nu}$ will nullify in equilibrium. This is a general phenomenon, applicable not only to self-sustained
systems but also to non-self-sustained systems such as weakly interacting
Bose-Einstein condensates discussed in~\cite{Finazzi2011}, and $^4$He and $^3$He solids. The nullification of the cosmological term is a consequence of the thermodynamic stability of the system,
 which implies an extremum of the function $\rho_{\rm vac}(c_{\perp})$, and of $\rho_{\rm vac}(g^{\mu\nu})$ in general. A non-zero but constant pressure,
such as the one necessary to maintain a gas in equilibrium, does therefore not affect the nullity
of the cosmological term.

Note that the nullification of the cosmological term corresponds to the experimental fact that quasiparticles in $^3$He-A near the zero-$T$ limit and phonons in a homogeneous sample of BEC move in straight trajectories, like free particles in a Minkowski spacetime. They certainly do not see a de Sitter or anti de Sitter spacetime with a huge cosmological constant, as could naively be expected based on the $(E_\text{Planck})^4$-dependence of the parameter $\lambda$.

\section{Cosmological constant and vacuum readjustment}
%{\em Cosmological constant and vacuum readjustment --}
Comparison of Eq.~\eqref{VacuumEnergyNon-equilibrium3} with the standard
form $\Lambda \sqrt{-g}$ does not allow us to unambiguously define the
cosmological constant $\Lambda$.
Nevertheless, a natural choice would be to use for the cosmological constant the
parameter $\lambda$ in  Eq.(\ref{VacuumEnergyNon-equilibrium2}),
$\Lambda=\lambda$. Incidentally, as follows from Eq.(\ref{VacuumEnergyNon-equilibrium2}),
this is the energy density of the original non-superfluid or false
vacuum state. Since $c_{\perp 0} \propto
\Delta_0 $, the false vacuum has  $c_{\perp}=0$,
$1/\sqrt{-g}=0$, and thus 
\begin{equation}
\rho_{\rm vac}({\rm false~vacuum})=\lambda  \sqrt{-g^{(0)}} \,.
\label{FalseVacuum}
\end{equation}
The false vacuum is an unstable equilibrium state corresponding to a local energy maximum in
the $T \to 0$ limit, which existed before the phase transition to the  (stable) equilibrium state 
(the true vacuum) with emerging gravity.  Eq.(\ref{FalseVacuum}) for the energy of the false vacuum is written assuming that the true vacuum exists in the absence of external environment, and thus $ \rho_{\rm vac}({\rm true~vacuum})=0$.  In the general case the cosmological constant $\lambda$ determines
the difference in the energies of the false and true vacua,  
\begin{equation}
\lambda  \sqrt{-g^{(0)}} = \rho_{\rm vac}({\rm false~vacuum}) - \rho_{\rm vac}({\rm true~vacuum})
 \,.
\label{FalseVacuum2}
\end{equation}
For example, if the false vacuum is originally in its quasi-equilibrium state in the absence of external environment, then its vacuum energy density is zero, $ \rho_{\rm vac}({\rm false~vacuum}) =0$, 
while the energy of the true vacuum is negative, $ \rho_{\rm vac}({\rm true~vacuum})=-\lambda  \sqrt{-g^{(0)}}$. After the phase transition to the superfluid state, the microscopic degrees of freedom are readjusted to the new equilibrium state, such that the energy of the true vacuum is nullified, while the energy of the false vacuum becomes positive~\cite{Volovik2006}.  

The identification of the parameter $\lambda$ with the cosmological constant becomes more explicit through its relation with the graviton masses in $^3$He-A. It is instructive to first recall some general results from relativistic bi-metric theories of gravity.

\section{Bi-metric relativistic theory of gravity}
%\label{RTG}
%{\em Bi-metric relativistic theory of gravity --}
Let us compare the bi-metric theory emerging in $^3$He-A with the so-called relativistic theory of
gravity (RTG) discussed in Ref.~\cite{LogunovMestvirishvili1985}. 

In~\cite{LogunovMestvirishvili1985}, the gravitational field is regarded as a physical field $\Phi^{\mu\nu}$ on top of a fundamental Minkowski background spacetime, which creates a curvature of the secondary, effective Riemannian spacetime. Regardless of the observational and conceptual issues of this theory\footnote{One generic problem with modifications of GR that are bi-metric and/or include massive gravitons is the possible presence of ghosts~\cite{Boulware:1973my}. In~\cite{Gershtein:2008kn} it is argued that such negative-energy contributions do not appear in RTG because the causality condition must necessarily be imposed in the (physical) background spacetime; see also~\cite{Pitts:2005wx} for several useful comments. As has been shown recently, it is possible to explicitly construct ghost-free models of massive gravity~\cite{Hassan:2011hr} and extend these to bi-metric gravity~\cite{Hassan:2011zd}. A rigorous study of the subtle relation between (ghost-free) bi-metric and massive models of gravity is given in~\cite{Baccetti:2012bk}, which also emphasises that there is no formal reason to consider only flat background metrics. We stress that we mention here Logunov's RTG because it provides the most direct route to the case relevant for our discussion of 
the effective gravity in $^3$He-A, namely that of a flat fundamental background metric, without committing ourselves to the underlying philosophy of RTG---or bi-metric theories in general---as opposed to GR.}, we are interested in the formal treatment of any free parameters in the theory which extend the concept of Einstein's cosmological constant. Due to the bi-metricity, one can in principle add two ``cosmological constants'' $\lambda_0\sqrt{-g^{(0)}}$ and $\Lambda \sqrt{-g}$, as well as two metric-coupling terms $\frac{1}{2}m^2 \sqrt{-g}g_{\mu\nu}^{(0)}g^{\mu\nu}$ and $\frac{1}{2} M^2 \sqrt{-g^{(0)}}g_{\mu\nu}g^{\mu\nu(0)}$ to the overall action 
without affecting the interplay between matter and the gravitational field. The parameters $\lambda_0$, $\Lambda$, $m$ and $M$ are constrained by 
the conservation equation $\nabla_\mu^{(0)} T^{\mu\nu}=0$ with respect to the Minkowski background metric $g^{(0)}_{\mu\nu}$ and by the consistency of the Minkowski limit $L_{\rm matter}=0; \Phi^{\mu\nu}=0$. This turns out to give $M=0$ and $\Lambda=\lambda_0$ as well as $m^2=16\pi G_{\rm N}\Lambda$. Furthermore, linearizing the global equations of motion gives $(\square - m^2)\Phi_{\mu\nu}=0$, hence $m$ corresponds to the graviton mass, which is equal for the usual spin $S=2$ graviton modes and the additional $S=0$ graviton\footnote{The bi-metric gravity with arbitrary mass ratio $\zeta$ can be found in~\cite{Babak2003}.}, and is thus directly related to the cosmological constant $\Lambda$.\footnote{The relation between the introduction of a graviton mass and a cosmological constant is actually a rather generic feature of many models of massive and bi-metric theories of gravity, although not necessarily in the simplest form that we discuss here, see~\cite{Baccetti:2012bk}.}

The bi-metric generalization of the ``vacuum energy'' term in the Einstein equations can then be written in the following way:
\begin{eqnarray} 
%\rho_{\rm vac}= 
 \rho_\lambda= \Lambda\left( \frac{1}{2}   \sqrt{-g}g^{(0)}_{\mu\nu}g^{\mu\nu}
-\sqrt{-g} -  \sqrt{-g^{(0)}} \right)=
\label{LambdaTerm1}
\\
\Lambda\left( \sqrt{-g} -  \sqrt{-g^{(0)}} + \frac{1}{2}  
\sqrt{-g}(g^{\mu\nu}-g^{(0)\mu\nu})
g^{(0)}_{\mu\nu}
\right)\,.
\label{LambdaTerm2}
\end{eqnarray}
%where $g^{(0)}_{\mu\nu}$ is the Minkowski background metric. 
 In the Minkowski vacuum, i.e. for $g_{\mu\nu}=g^{(0)}_{\mu\nu}$, which is an
analogue of the
 equilibrium vacuum in $^3$He-A, $\rho_\lambda$ is zero. %the vacuum energy is zero. 
The cosmological term 
\begin{align} 
 T_{\mu\nu}^\Lambda &\equiv
\frac{2}{\sqrt{-g}}\frac{\partial  \rho_\lambda}
{\partial g^{\mu\nu}} \nonumber\\ &= 
-\Lambda\left
(\left(g_{\mu\nu}-g^{(0)}_{\mu\nu}\right)  
+ \frac{1}{2}g_{\mu\nu} g^{(0)}
_{\alpha\beta}\left(g^{\alpha\beta}-g^{(0)\alpha\beta}\right) 
\right)\,,
\label{CosmologicalTerm}
\end{align}
 is also zero in the Minkowski vacuum,  even though the cosmological constant
$\Lambda \propto m^2G_N^{-1}$ itself is nonzero and is on the order of the Planck scale.

\section{Cosmological constant and massive gravitons}
%{\em Cosmological constant and massive gravitons --}
Eq.(\ref{VacuumEnergyNon-equilibrium2}) is the vacuum energy in $^3$He-A as a
function of $c_{\perp}$ only. Let us consider the more general energy as a function of the
metric elements $g^{11}$,  $g^{22}$ and $g^{12}$, which are induced in $^3$He-A, and compare with the RTG of the previous Section.
Expressing the perturbations of the metric in terms of the variables $\eta_i$
corresponding to propagating gravitons
\begin{gather}
\delta g^{11} =c_{\perp 0}^2  \frac{\eta_0+\eta_1}{\sqrt{2}}\,,~~~
\delta g^{22} =c_{\perp 0}^2  \frac{\eta_0-\eta_1}{\sqrt{2}}\,,\\
\delta g^{12} =\delta g^{21}=c_{\perp 0}^2  \frac{\eta_2}{\sqrt{2}}
\,,
\label{Elements}
\end{gather}
one obtains ~\cite{Volovik1986b} the non-equilibrium vacuum density
\begin{equation}
%\rho_{\rm vac}
  \rho_{\rm vac}^{\rm pert}
=\frac{1}{4}\lambda  \sqrt{-g^{(0)}}\left(\eta_0^2 +\frac{1}{2}( \eta_1^2 +
\eta_2^2 )\right) \,.
\label{Action}
\end{equation}
For $g^{11}=g^{22}=c_{\perp}^2$ and $g^{12}=0$, this transforms to
Eq.(\ref{VacuumEnergyNon-equilibrium3}). 

The   so-called clapping modes
 $\eta_1$ and $\eta_2$ are equivalent to gravitons propagating along the
anisotropy axis with  spin 
(helicity) $S=2$.  The symmetry between these two $S=2$ graviton modes
means that their coefficients in Eq.~\eqref{Action} must necessarily be equal.
In addition there is a mode $\eta_0$ which is an analogue of
the propagating graviton with spin $S=0$ (for this mode the analogy with
relativistic theories is not perfect:  in $^3$He-A the perturbation of the
metric is not traceless in the $S=0$ mode since the perturbation of the metric
element $g^{33}$ is missing). 

One can write  the $^3$He-A energy in Eq.(\ref{Action}) in a manner similar to $\rho_\Lambda$ in Eq.\eqref{LambdaTerm2} for the
RTG, as an expansion of the following expression:
\begin{equation} 
  \rho_{\rm vac}=
 \frac{\lambda}{2}\left( \sqrt{-g} -  \sqrt{-g^{(0)}} + \frac{1}{2}  
\sqrt{-g^{(0)}}(g^{\mu\nu}-g^{(0)\mu\nu})
g^{(0)}_{\mu\nu}
\right)\,.
\label{LambdaTermHe3}
\end{equation}

The energy (\ref{Action}) determines the masses $m_i$ of the gravitons, which are seen to obey the ratio
\begin{equation}
\zeta=\frac{m^2_{S=0}}{m^2_{S=2}}=2 \,.
\label{MassRatio}
\end{equation} 
In relativistic bi-metric theories, such as the RTG discussed previously, the mass of the gravitons is determined by the 
product of the cosmological constant and the Newton constant, $m^2 \sim
G_{\rm N}\lambda$. In $^3$He-A, the Newton constant is not well defined, because due to the lack of
general covariance the terms in the action which
describe gravity are not combined  into a single Einstein action, and so they have different ``Newton constants'' ~\cite{Volovik2003,Volovik1998}. This can clearly be seen in the
example of the gravitons propagating along the anisotropy axis. 
The effective action for these gravitons explicitly contains the mass term \eqref{Action} (or \eqref{LambdaTermHe3})~\cite{Volovik1986b}, and it is an easy exercise to derive their 
equations of propagation. These contain two ``Newton constants'', $G_{\rm N}$ and $\tilde G_{\rm N}$, which enter 
correspondingly to the space derivative and time derivative sectors of the Einstein tensor:
\begin{align}
 \frac{1}{16\pi \tilde G_{\rm N}} \partial_t^2\eta_{1,2} & - \frac{c_\parallel^2}{16\pi G_{\rm N}} \partial_z^2\eta_{1,2} + \lambda\eta_{1,2}=0 \,,
  \label{NewtonConstants1}
\\
  \frac{1}{16\pi \tilde G_{\rm N}} \partial_t^2\eta_{0} & - \frac{c_\parallel^2}{16\pi G_{\rm N}} \partial_z^2\eta_{0} + \zeta\lambda\eta_{0}=0\,.
 \label{NewtonConstants2}
\end{align}
The Newton constants are on the order of the Planck scale,
$G_{\rm N}\sim \Delta_0^{-2}$. In the BCS limit they obey the relation $G_{\rm N}=3\tilde G_{\rm N}$, while the masses of the gravitons are $m^2_{S=0}=(8/3)\Delta_0^2$ and $m^2_{S=2}=(4/3)\Delta_0^2$. 
This difference in mass ratio ($\zeta=2$ in Eq. (\ref{MassRatio}) in $^3$He-A, as compared with
$\zeta=1$ in the RTG) is caused by the minimal difference between RTG and the bi-metric gravity in $^3$He-A ($\sqrt{-g^{(0)}}$
in the third term of Eq.\eqref{LambdaTermHe3} instead of $ \sqrt{-g}$ in Eq.(\ref{LambdaTerm2})), as well as the different composition of the $S=0$ graviton.\footnote{In the future it would be interesting to discuss whether the problem of the van Dam-Veltman-Zakharov discontinuity~\cite{vanDam1970,Zakharov1970} can be resolved in the effective gravity emerging in the systems with Weyl points.}
  But crucially, Eqs. \eqref{NewtonConstants1}-\eqref{NewtonConstants2} show that the link between the graviton masses and the generalized bi-metric cosmological constant $\lambda$ in RTG survives in $^3$He-A. Moreover, in $^3$He-A, this term is directly determined by the vacuum energy through Eqs.\eqref{VacuumEnergyNon-equilibrium2} and \eqref{FalseVacuum2}.

Finally, comparison of Eqs.~\eqref{LambdaTermHe3} and \eqref{LambdaTerm2} shows that, even though the action in $^3$He-A is not generally covariant, it makes good sense to define a cosmological term in analogy with \eqref{CosmologicalTerm}:
\begin{equation}
 T_{\mu\nu}^\Lambda=\frac{2}{\sqrt{-g}}\frac{\partial \rho_{\rm vac}}{\partial g^{\mu\nu}}=-\frac{\lambda}{\sqrt{-g}}\left(\sqrt{-g}g_{\mu\nu} - \sqrt{-g^{(0)}}g_{\mu\nu}^{(0)}\right)
\end{equation}
which indeed vanishes in equilibrium, as required by the thermodynamic arguments discussed above.

\section{Grand canonical vacuum energy and cosmological constant}
%{\em Grand canonical vacuum energy and cosmological constant --}
%
Using their example of effective gravity in BEC, the authors of Ref.
\cite{Finazzi2011} state that ``it is conceivable that the very notion of
cosmological constant as a form of energy intrinsic
to the vacuum is ultimately misleading''.  From our point of view this
conclusion is based on an example where the effective gravity is very far from
GR or from a relativistic bi-metric theory of gravity. In our example, gravity in $^3$He-A  is
a bi-metric gravity. Though this gravity emerges on a  particular non-relativistic
background, its characteristics are universally determined by the topological class (the Weyl point).
The effective theory thus obeys both Lorentz invariance and the equivalence principle, and
some terms in the effective action for gravity are similar to that
in relativistic bi-metric gravity, which in particular describe the propagation
of gravitons. The latter allows us to identify the analogue of the cosmological
constant.  In $^3$He-A, this cosmological constant is directly related to the properly defined vacuum energy.
The cosmological constant turns out to be equal to the difference in energy
densities of the vacuum states with and without gravity. In other words, the
cosmological constant coincides with that part of the vacuum energy which comes
from the degrees of freedom responsible for the emergence of the effective
low-energy space-time and thus for the effective gravity. 
 Moreover, the close relationship with a relativistic bi-metric theory suggests to maintain the definition of the cosmological term $T^\Lambda_{\mu\nu}$, which vanishes in equilibrium.
Contrarily to GR, for which the high-energy theory is unknown, in the effective gravity of  $^3$He-A, the microscopic physics can be traced explicitly. $^3$He-A thus shows a possible scenario for the decoupling of the high-energy degrees of freedom, which could be relevant for GR as well.

A related example is provided by  the $q$-theory -- the phenomenological theory
of the quantum vacuum 
\cite{KlinkhamerVolovik2008,KlinkhamerVolovik2008b,KlinkhamerVolovik2011a}. 
In this theory the quantity $q$ is the Lorentz invariant variable, which
characterizes the vacuum. At first glance, the connection between the
cosmological constant $\Lambda$ and the vacuum energy is lost  in the $q$
theory: the cosmological constant is not equal to the vacuum energy
$\epsilon(q)$ of the field $q$. But it coincides with the analog of the grand
canonical energy of the field, $\Lambda=\rho_{\rm vac}(q) =\epsilon(q) -q
d\epsilon/dq$: from a thermodynamic analysis and from the dynamic equations of
the $q$-theory it follows that it is this energy $\rho_{\rm vac}(q)$ which is
gravitating, rather than $\epsilon(q)$. In this $q$-theory, cosmology is the
process of relaxation of the Universe towards the equilibrium vacuum state. The
vacuum energy density $\rho_{\rm vac}$ has initially a Planck-scale value, which
corresponds to the $^3$He false vacuum state without gravity. In the process of
relaxation,  $\rho_{\rm vac}$ drops to zero value in the final equilibrium
vacuum, while the energy $\epsilon(q)$ is still of the Planck scale value.  The dynamics of this relaxation process depends on the precise microscopic theory and requires the detailed consideration
(see \cite{EmelyanovKlinkhamer2012}), but its final equilibrium state always corresponds to the Minkowski spacetime. 
This property of the final state of the Universe is therefore similar to the property  of
equilibrium $^3$He-A.  
\footnote{A small remnant cosmological constant in our present Universe may result from the contribution of the neutrino sector to the vacuum energy~\cite{KlinkhamerVolovik2011a}. Each fermionic field has its own contribution, which is  the negative energy of the corresponding Dirac vacuum, while in the equilibrium vacuum all these contributions are compensated by the microscopic degrees of freedom -- the $q$-type field. When the fermionic field acquires mass its negative contribution to the vacuum energy increases. This corresponds to a decrease of the overall positive vacuum energy, which is relaxed due to the fast oscillations of the $q$-field. In the case of the neutrinos, the relaxation time of the vacuum could be large, so that the negative mass-effect contribution $\rho_{\rm vac}^\nu \sim -M_\nu^4$ is still  missing. The lack of a negative neutrino contribution to the vacuum energy density corresponds, of course, to a positive remnant cosmological constant $\Lambda \sim M_\nu^4$.}

 \section{Conclusion}
%{\em Conclusion --}
In conclusion, the notion of cosmological constant as a form of energy intrinsic
to the vacuum is not misleading. The  closer the emergent gravity is to GR, the
better the connection between these two notions. The nonrelativistic systems with Weyl fermions
and relativistic $q$-theory teach us that the cosmological constant is related
to the  grand canonical energy of the vacuum, which is relevant for
thermodynamics. The grand canonical energy is the proper vacuum energy, which
excludes the irrelevant  contributions of high-energy degrees of freedom from
the cosmological constant. The effective gravity emerging in nonrelativistic
$^3$He-A is a bi-metric gravity, where the cosmological term is complicated and
the notion of the cosmological constant  is ambiguous. Nevertheless, the
parameter $\lambda$ which enters the cosmological term in
Eq.(\ref{VacuumEnergyNon-equilibrium2})  is also the one that determines the graviton masses, and this explicitly establishes that it is the natural bi-metric generalization of the
cosmological constant. This cosmological constant has a definite relation to the vacuum energy. It is the
difference in the proper energy of the vacuum in two states: the nonequilibrium
state without gravity and the equilibrium state in which gravity emerges.  This
difference is of the Planck energy scale, which however does not pose a problem
for cosmology: the cosmological term vanishes in the equilibrium vacuum, as dictated by thermodynamics. This suggests that the final state of the Universe is not gravitating, independently of the microscopic details of the relaxation process through which this final state is reached.

\subsection*{Acknowledgements}
The authors thank Carlos Barcel\'{o} and Luis J.~Garay for useful comments. This work is supported in part by the Academy of Finland and its COE program, with a contribution from the EU 7th Framework Programme (FP7/2007-2013,  grant 228464 Microkelvin). G.J. also acknowledges a FECYT postdoctoral mobility contract
 of the Spanish MEC/MICINN.


\vspace*{3ex}
\begin{thebibliography}{99}

\bibitem{Finazzi2011} 
S. Finazzi, S. Liberati and L. Sindoni,
The cosmological constant: a lesson from Bose-Einstein condensates,
Phys. Rev. Lett. {\bf 108}, 071101 (2012);
arXiv:1103.4841.


\bibitem{Sindoni2011} 
L. Sindoni,
Emergent Models for Gravity: an Overview of Microscopic Models.
SIGMA {\bf 8}, 027 (2012).
arXiv:1110.0686.

\bibitem{Barcelo:2005fc} 
  C.~Barcel\'o, S.~Liberati and M.~Visser,
  Analogue gravity,
  Living Rev.\ Rel.\  {\bf 8}, 12 (2005).

\bibitem{Volovik2003}
G.E. Volovik,
The Universe in a Helium Droplet, 
Clarendon Press,  Oxford (2003).

\bibitem{Abrikosov1971}
A.A. Abrikosov    and S.D. Beneslavskii,
Possible existence of substances intermediate between metals and dielectrics, 
Sov. Phys. JETP {\bf 32}, 699 (1971). 

 \bibitem{Abrikosov1998}
A.A. Abrikosov,
 Quantum magnetoresistance,  Phys. Rev. {\bf B 58}, 2788 (1998).

\bibitem{XiangangWan2011}
Xiangang Wan, A.M. Turner,  A. Vishwanath  and S.Y. Savrasov,
 Topological semimetal and Fermi-arc surface states in the electronic structure
of pyrochlore iridates,
Phys. Rev. B {\bf 83}, 205101 (2011).

\bibitem{Burkov2011}
A.A. Burkov and L. Balents, 
Weyl semimetal in a topological insulator multilayer,
Phys. Rev. Lett. {\bf 107}, 127205 (2011);
arXiv:1105.5138.

 \bibitem{Aji2011}
 Vivek Aji,
Adler-Bell-Jackiw anomaly in Weyl semi-metals: Application to Pyrochlore
Iridates,
arXiv:1108.4426.

\bibitem{Vozmediano2010} 
M. A. H. Vozmediano, M. I. Katsnelson, F. Guinea,
Gauge fields in graphene,
Physics Reports {\bf 496}, 109 (2010).

\bibitem{CveticGibbons2012}
M. Cvetic, G.W Gibbons,
Graphene and the Zermelo Optical Metric of the BTZ Black Hole,
arXiv:1202.2938

\bibitem{Horava2005}  
P. Ho\v{r}ava,
Stability of Fermi surfaces and $K$-theory,
Phys. Rev. Lett. \textbf{95}, 016405 (2005).

\bibitem{Stone2012} 
M. Stone, 
Gravitational anomalies and thermal Hall effect in topological Insulators,
Phys. Rev. B {\bf 85}, 184503 (2012);
arXiv:1201.4095.

\bibitem{Griffin2012} 
T. Griffin, P. Ho\v{r}ava and C.M. Melby-Thompson,
Conformal Lifshitz gravity from holography,
JHEP {\bf 1205}, 010 (2012)
  [arXiv:1112.5660 [hep-th]].
  %%CITATION = ARXIV:1112.5660;%%

\bibitem{KatsnelsonVolovik2012}
M.I. Katsnelson and G.E. Volovik,
Quantum electrodynamics with anisotropic scaling: Heisenberg-Euler action and Schwinger pair production in the bilayer graphene,
  Pisma Zh.\ Eksp.\ Teor.\ Fiz.\  {\bf 95}, 457 (2012)
  [arXiv:1203.1578 [cond-mat.str-el]].
  %%CITATION = ARXIV:1203.1578;%%
  
\bibitem{Zubkov2012}
M.A. Zubkov,
Schwinger pair creation in multilayer graphene
Pis'ma ZhETF {\bf 95}, 540 (2012); arXiv:1204.0138.

\bibitem{einstein}
A.~Einstein,
Autobiographical notes. In: P.A. Schilpp (ed.), ``Albert Einstein: Philosopher-Scientist'', Cambridge Univ. Pr. (1949).

\bibitem{jacobson}
  T.~Jacobson,
  Thermodynamics of space-time: The Einstein equation of state,
  Phys.\ Rev.\ Lett.\  {\bf 75}, 1260 (1995)
  [gr-qc/9504004].
  %%CITATION = GR-QC/9504004;%%

\bibitem{padmanabhan}
  T.~Padmanabhan,
  ``Thermodynamical Aspects of Gravity: New insights,''
  Rept.\ Prog.\ Phys.\  {\bf 73}, 046901 (2010)
  [arXiv:0911.5004 [gr-qc]].
  %%CITATION = ARXIV:0911.5004;%%

\bibitem{blhu}
  B.~L.~Hu,
  ``Gravity and Nonequilibrium Thermodynamics of Classical Matter,''
  Int.\ J.\ Mod.\ Phys.\ D {\bf 20}, 697 (2011)
  [arXiv:1010.5837 [gr-qc]].
  %%CITATION = ARXIV:1010.5837;%%


\bibitem{Volovik2006} 
G.E. Volovik,
Vacuum Energy: Myths and Reality,  
Int. J. Mod. Phys. {\bf D15},   1987--2010 (2006); 
  gr-qc/0604062.

\bibitem{KlinkhamerVolovik2008}
 F.R. Klinkhamer and G.E. Volovik,  
 Self-tuning vacuum variable and cosmological constant, 
  Phys. Rev. D {\bf 77}, 085015 (2008).

 \bibitem{KlinkhamerVolovik2008b}
 F.R. Klinkhamer and G.E. Volovik,  
 Dynamic vacuum variable and equilibrium approach in cosmology,   
 Phys. Rev. D {\bf 78}, 063528 (2008).
 
\bibitem{KlinkhamerVolovik2011a} 
  F.R. Klinkhamer and G.E.Volovik,
Dynamics of the quantum vacuum: Cosmology as relaxation to the equilibrium
state,
 J. Phys. Conf. Ser. {\bf 314}, 012004 (2011); 
 arXiv:1102.3152.


\bibitem{Liberati:2001sd} 
  S.~Liberati, S.~Sonego and M.~Visser,
  Faster than c signals, special relativity, and causality,
  Annals Phys.\  {\bf 298}, 167 (2002)

\bibitem{Barcelo:2007iu} 
  C.~Barcel\'o and G.~Jannes,
  A Real Lorentz-FitzGerald contraction,
  Found.\ Phys.\  {\bf 38}, 191 (2008)


\bibitem{atiyah} 
M.F. Atiyah, R. Bott and A. Shapiro, 
Clifford Modules,
Topology 3 Suppl. {\bf 1}, 3 (1964).

\bibitem{Froggatt:1991ft} 
  C.~D.~Froggatt and H.~B.~Nielsen,
  ``Origin of symmetries,''
  World Scientific,   Singapore (1991).

\bibitem{Volovik:2006gt} 
  G.~E.~Volovik,
  ``Quantum phase transitions from topology in momentum space,''
  Lect.\ Notes Phys.\  {\bf 718}, 31 (2007)
 [cond-mat/0601372 [cond-mat.str-el]].
  %%CITATION = COND-MAT/0601372;%%

\bibitem{LogunovMestvirishvili1985}
A. A. Logunov and M. A. Mestvirishvili,
Relativistic theory of gravitation and the graviton rest mass,
Teoret. Mat. Fiz. {\bf 65}, 3--15 (1985);  Theoretical and Mathematical Physics
{\bf 65}, 971--979 (1985).

%\cite{Boulware:1973my}
\bibitem{Boulware:1973my} 
  D.~G.~Boulware and S.~Deser,
  ``Can gravitation have a finite range?,''
  Phys.\ Rev.\ D {\bf 6}, 3368 (1972).
  %%CITATION = PHRVA,D6,3368;%%

%\cite{Gershtein:2008kn}
\bibitem{Gershtein:2008kn} 
  S.~S.~Gershtein, A.~A.~Logunov, M.~A.~Mestvirishvili and ,
  ``Gravitational Waves in Relativistic Theory of Gravitation,''
  Theor.\ Math.\ Phys.\  {\bf 160}, 1096 (2009)
  [arXiv:0810.4393 [gr-qc]].
  %%CITATION = ARXIV:0810.4393;%%

%\cite{Pitts:2005wx}
\bibitem{Pitts:2005wx} 
  J.~B.~Pitts and W.~C.~Schieve,
  ``Universally coupled massive gravity''
  Theor.\ Math.\ Phys.\  {\bf 151}, 700 (2007)
  [gr-qc/0503051].
  %%CITATION = GR-QC/0503051;%%

%\cite{Hassan:2011hr}
\bibitem{Hassan:2011hr} 
  S.~F.~Hassan and R.~A.~Rosen,
  ``Resolving the Ghost Problem in non-Linear Massive Gravity,''
  Phys.\ Rev.\ Lett.\  {\bf 108}, 041101 (2012)
  [arXiv:1106.3344 [hep-th]].
  %%CITATION = ARXIV:1106.3344;%%

%\cite{Hassan:2011zd}
\bibitem{Hassan:2011zd} 
  S.~F.~Hassan and R.~A.~Rosen,
  ``Bimetric Gravity from Ghost-free Massive Gravity,''
  JHEP {\bf 1202}, 126 (2012)
  [arXiv:1109.3515 [hep-th]].
  %%CITATION = ARXIV:1109.3515;%%


%\cite{Baccetti:2012bk}
\bibitem{Baccetti:2012bk} 
  V.~Baccetti, P.~Martin-Moruno and M.~Visser,
  ``Massive gravity from bimetric gravity,''
  arXiv:1205.2158 [gr-qc].
  %%CITATION = ARXIV:1205.2158;%%

\bibitem{Babak2003}
S.V. Babak and L.P. Grishchuk,
Finite-range gravity and its role in gravitational waves, black holes and
cosmology,
Int. J. Mod. Phys. D{\bf 12}, 1905--1960 (2003).


\bibitem{Volovik1986b} 
G.E. Volovik, 
Analog of gravity in superfluid  $^3$He-A,
JETP Lett. {\bf 44}, 498--501 (1986).

\bibitem{Volovik1998} 
G.E. Volovik, 
Gravity of monopole and string and gravitational constant in $^3$He-A,   
JETP Lett. {\bf 67}, 698 - 704  (1998); cond-mat/9804078. 

\bibitem{vanDam1970} 
 H. van Dam and M.G. Veltman,
 Massive and massless Yang-Mills and gravitational field,
 Nucl. Phys. B {\bf 22}, 397--411 (1970).

\bibitem{Zakharov1970}  
V. I. Zakharov, 
Linearized gravitation theory and the graviton mass,
JETP Lett. {\bf 12}, 312--314 (1970).

\bibitem{EmelyanovKlinkhamer2012}  
  V.~Emelyanov and F.~R.~Klinkhamer,
  ``Possible solution to the main cosmological constant problem,''
  Phys.\ Rev.\ D {\bf 85}, 103508 (2012)
  [arXiv:1109.4915 [hep-th]].
  %%CITATION = ARXIV:1109.4915;%%

\end{thebibliography}
\end{document}